\documentclass[aps,twocolumn,groupedaddress,amsmath,amssymb]{revtex4}

\usepackage{graphicx}
\usepackage{dcolumn}
\usepackage{xcolor}%
\usepackage{booktabs}%
\usepackage{bm}

\begin{document}

\title{Enhancement of superconductivity on the verge of a structural instability in isovalently doped $\beta$-ThRh$_{1-x}$Ir$_{x}$Ge}

\date{\today}
\author{Guorui Xiao$^{1,2,3}$}
\email{xiaoguorui@westlake.edu.cn}
\author{Qinqing Zhu$^{1,2,4}$}
\author{Yanwei Cui$^{1,2,3}$}
\author{Wuzhang Yang$^{1,2,4}$}
\author{Baizhuo Li$^{3}$}
\author{Shijie Song$^{3}$}
\author{Guang-Han Cao$^{3}$}
\author{Zhi Ren$^{1,2}$}
\email{renzhi@westlake.edu.cn}

\affiliation{$^{1}$School of Science, Westlake University, 18 Shilongshan Road, Hangzhou, 310024, Zhejiang Province, PR China}
\affiliation{$^{2}$Institute of Natural Sciences, Westlake Institute for Advanced Study, 18 Shilongshan Road, Hangzhou, 310024, Zhejiang Province, PR China}
\affiliation{$^{3}$Department of Physics, Zhejiang University, Hangzhou 310027, P. R. China}
\affiliation{$^{4}$Department of Physics, Fudan University, Shanghai, 200433, PR China}

\begin{abstract}
$\beta$-ThRhGe, the high-temperature polymorph of ThRhGe, is isostructural to the well-known ferromagnetic superconductor URhGe. However, contrary to URhGe, $\beta$-ThRhGe is nonmagnetic and undergoes an incomplete structural phase transition at 244 K, followed by a superconducting transition below 3.36 K. Here we show that the isovalent substitution of Ir for Rh leads to a strong enhancement of superconductivity by suppressing the structural transition. At $x$ = 0.5, where the structural transition disappears, $T_{\rm c}$ reaches a maximum of 6.88 K. The enhancement of superconductivity is linked to the proximity to a structural quantum critical point at this Ir concentration, as suggested by the analysis of thermodynamic as well as resistivity data. First principles calculations indicate that the Ir doping has little effect on the electronic band dispersion near the Fermi level. $\beta$-ThRh$_{1-x}$Ir$_{x}$Ge thus provides an excellent platform to study the interplay between superconductivity and structural quantum criticality in actinide-containing compounds.
\end{abstract}

\maketitle
\maketitle

\noindent\textbf{INTRODUCTION}\\
The interplay between structural instability and superconductivity has attracted sustainable attention over the past few decades. In particular, the superconducting transition temperature $T_{\rm c}$ is often enhanced as the  structural phase transition is suppressed by chemical doping or external pressure, which is of interest from both fundamental and application points of view. Thus far, such enhancement has been observed in a variety of systems, including elements \cite{PhysRevLett.44.1623,PhysRevLett.96.047003,PhysRevLett.77.1151}, intermetallic compounds \cite{RevModPhys.47.637,PhysRevLett.98.067002,PhysRevB.94.224508,song2020pressure,PhysRevB.86.100505,PhysRevLett.109.097002,hlukhyy2017structural,PhysRevLett.109.237008,PhysRevB.89.075117,
PhysRevLett.115.207003,PhysRevLett.114.097002,chen2019superconductivity,PhysRevB.87.121107,PhysRevB.93.140505}, transition metal dichalcogenides \cite{qi2016superconductivity,PhysRevB.95.100501}, cuprates \cite{lee2006interplay,reznik2006electron}, and iron pnictides \cite{kamihara2008iron,de2008magnetic,PhysRevLett.103.057002,paglione2010high}. In several cases, the disappearance of structural transition coincides with the maximum in $T_{\rm c}$, which is taken as evidence for a structural quantum critical point (QCP) \cite{PhysRevLett.115.207003,PhysRevLett.114.097002,gruner2017charge}. However, in actinide-containing compounds, where many exotic properties have been found, no such example has been reported to date.

\begin{figure*}
	\includegraphics*[width=17.5cm]{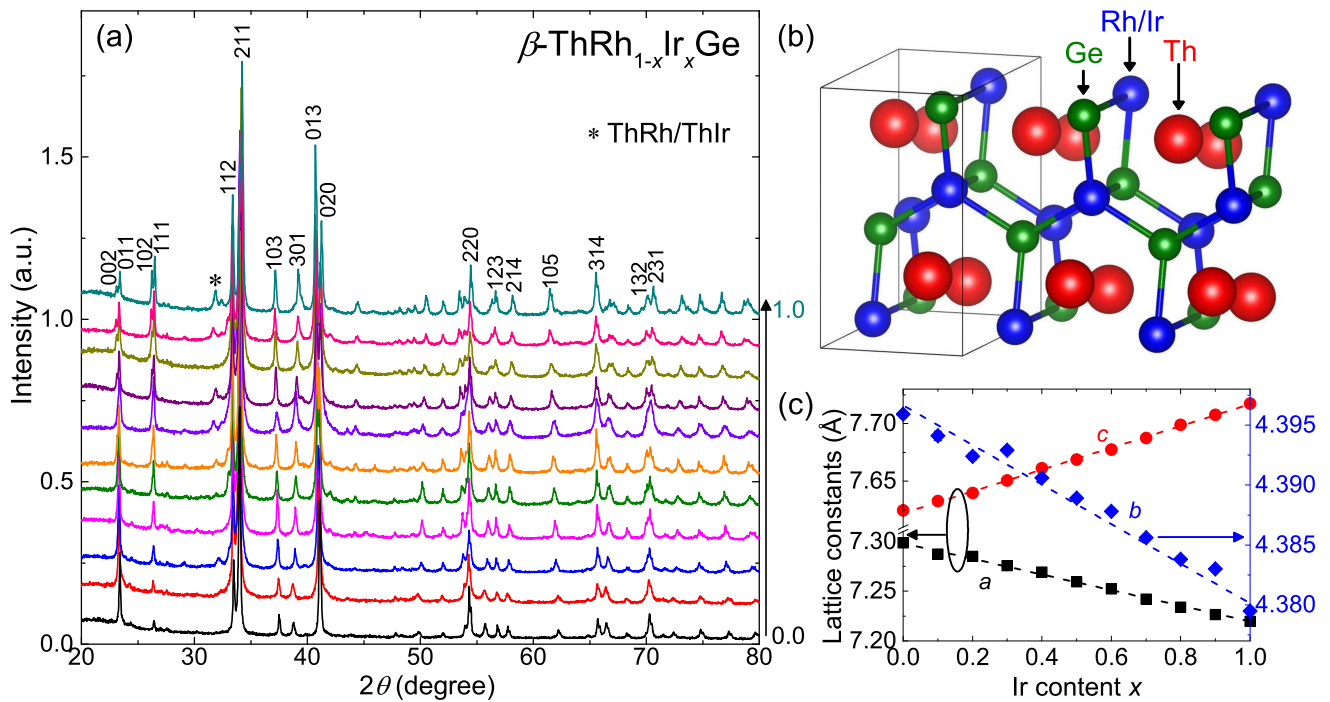}
	\caption{(Color online) \textbf{X-ray diffraction results and schematic crystal structure of $\beta$-ThRh$_{1-x}$Ir$_{x}$Ge.}
		(a) Room temperature XRD patterns for the series of polycrystalline $\beta$-ThRh$_{1-x}$Ir$_{x}$Ge samples. The diffraction peak due to the ThRh/ThIr impurity is marked by the asterisk.
        (b) Schematic structure of $\beta$-ThRh$_{1-x}$Ir$_{x}$Ge. The Th, Rh/Ir and Ge atoms are marked by the arrows.
		(c) Ir content $x$ dependence of the orthorhombic lattice constants.
	}
	\label{fig1}
\end{figure*}
Very recently, a new ternary equiatomic germanide ThRhGe has been synthesized and characterized \cite{xiao2022}. Depending on the annealing temperature, two orthorhombic polymorphs are obtained at ambient condition. $\alpha$-ThRhGe crystallizes in the YPdSi-type structure ($Pmmn$) and displays a normal metallic behavior down to 1.8 K. On the other hand, $\beta$-ThRhGe adopts the same TiNiSi-type structure ($Pnma$) as URhGe, which exhibits both ferromagnetism and superconductivity \cite{aoki2001coexistence}. By contrast, the nonmagnetic $\beta$-ThRhGe undergoes an incomplete transition into the monoclinic structure ($P2_{1}/c$) on cooling below 244 K and becomes superconducting below $T_{\rm c}$ = 3.36 K. 
The structural transition is of first order and accompanied by a bump in resistivity, a drop in magnetic susceptibility and a distinct specific heat anomaly.
Below the transition, the orthorhombic and monoclinic phases are found to coexist, though the precise atomic position in the latter remains to be determined when single crystalline samples become available.
Notably, the application of hydrostatic pressure suppresses the structural transition and enhances $T_{\rm c}$, whose onset reaches 8.36 K at 2.8 GPa.
To our knowledge, $\beta$-ThRhGe represents the first actinide-containing compound that shows the concurrence of structural transition and superconductivity, which provides an emerging platform to study the interplay between the two phenomena. In this respect, it is noted that the TiNiSi-type sister compound ThIrGe is also a superconductor with $T_{\rm c}$ = 5.25 K and exhibits no structural transition \cite{xiao2021synthesis}. Hence a systematic investigation of the isovalently doped $\beta$-ThRh$_{1-x}$Ir$_{x}$Ge series is worth pursuing.

Motivated by this, we present a systematic study on the structural and superconducting properties of $\beta$-ThRh$_{1-x}$Ir$_{x}$Ge across the whole $x$ range of 0 $\leq$ $x$ $\leq$ 1.
It is found that the disappearance of structural transition coincides with the maximum in $T_{\rm c}$ at $x$ = 0.5.
At this $x$ value, extremes in thermodynamic parameters as well as a non-Fermi liquid behavior are also observed, providing evidence for the existence of a structural QCP.
Furthermore, we show that Ir doping and application of hydrostatic pressure affect the structural transition and superconductivity in a very similar way.
The effect of Ir doping on the electronic band structure is investigated by theoretical calculations, whose results are discussed in comparison with the experimental observations.\\

\noindent\textbf{RESULTS AND DISCUSSION}\\
\noindent\textbf{XRD at room temperature}\\
\begin{figure}
	\includegraphics*[width=8cm]{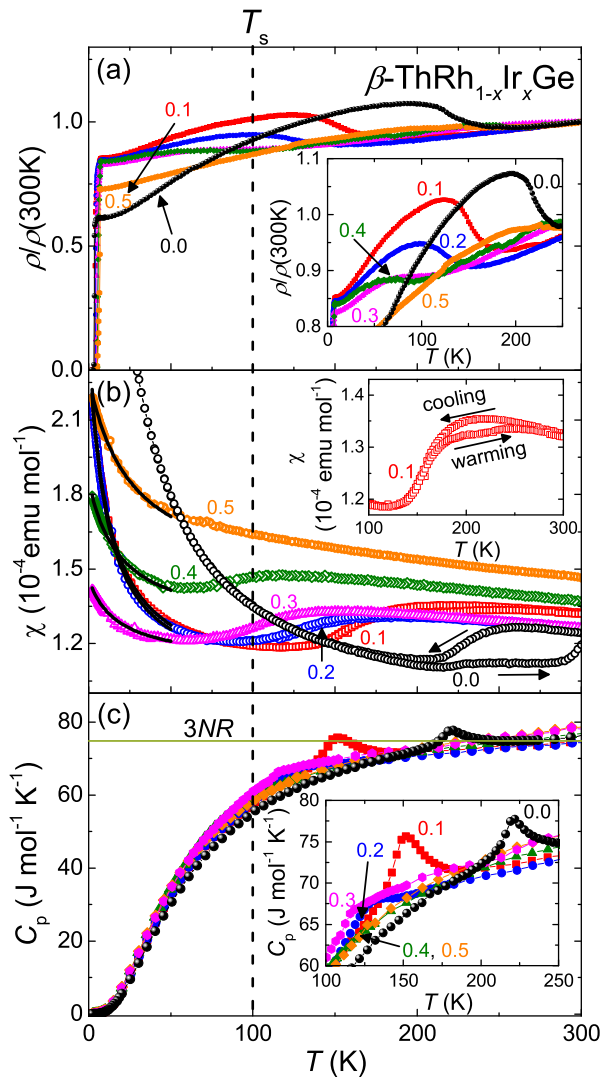}
	\caption{(Color online) \textbf{Evolution of physical properties in $\beta$-ThRh$_{1-x}$Ir$_{x}$Ge with $x$ $\leq$ 0.5.}
		(a-c) Temperature dependencies of resistivity, magnetic susceptibility and specific heat, respectively, for the $\beta$-ThRh$_{1-x}$Ir$_{x}$Ge samples with $x$ $\leq$ 0.5.
        The vertical dashed line indicates the $T_{\rm s}$ for $x$ = 0.4.
        In panel (a), the arrow marks the $x$ increasing direction and the inset shows a zoom of the data near the structural transition.
        In panel (b), the solid lines are fits to the data below 50 K by the Curie-Weiss law. The inset shows a zoom of the data for $x$ = 0.1 near the transition and the two arrows mark the temperature changing directions.
		In panel (c), the horizontal line corresponds to the Dulong-Petit limit and the inset shows a zoom of the data near the structural transition.
	}
	\label{fig2}
\end{figure}
Figure 1(a) shows the room temperature XRD patterns for the series of $\beta$-ThRh$_{1-x}$Ir$_{x}$Ge samples. For all $x$ values, the major diffraction peaks can be well indexed on the TiNiSi-type orthorhombic structure with the $Pnma$ space group, and a small amount of ThRh/ThIr impurity is also identified. A schematic structure of $\beta$-ThRh$_{1-x}$Ir$_{x}$Ge is displayed in Fig. 1(b). One can see that the Th atoms are located in the cavity of Rh/Ir-Ge three-dimensional network and form zigzag chains running along the $a$-axis. As a consequence, the inversion symmetry is absent for all atoms along the $c$-axis. The lattice parameters determined by the Lebail fitting are plotted as a function of Ir content $x$ in Fig. 1(c). With increasing $x$, the $a$- and $b$-axis shrink from 7.2872 {\AA} and 4.3941 {\AA} to 7.2266 {\AA} and 4.3830 {\AA}, respectively. This is somewhat unexpected since the atomic radius of Ir (1.355 {\AA}) is slightly larger than that of Rh (1.342 {\AA}) \cite{pauling1947atomic}. On the contrary, the $c$-axis increases from 7.6328 {\AA} to 7.7075 {\AA}. Despite this difference, it is noted that all the three axes vary almost linearly, in line with the Vegard's law. Hence the isovalent substitution of Ir for Rh in $\beta$-ThRh$_{1-x}$Ir$_{x}$Ge leads to not only the formation of a continuous solid solution but also an anisotropic change in the orthorhombic unit cell.\\

\noindent\textbf{Suppression of the structural phase transition}
\begin{figure*}
	\includegraphics*[width=16.8cm]{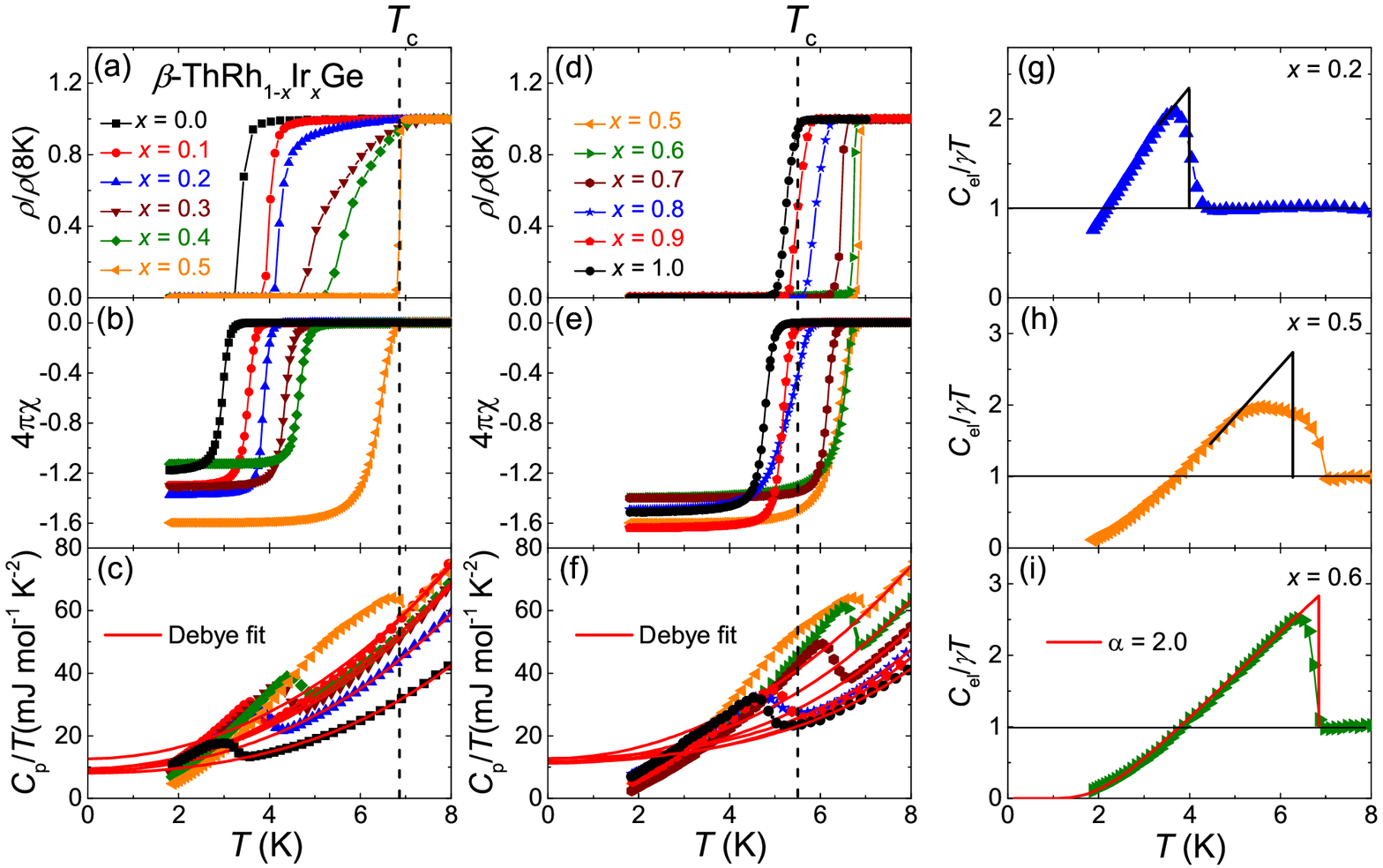}
	\caption{(Color online) \textbf{Evolution of superconductivity in $\beta$-ThRh$_{1-x}$Ir$_{x}$Ge.}
		(a-c) Low-temperature resistivity, magnetic susceptibility and specific heat, respectively, for the $\beta$-ThRh$_{1-x}$Ir$_{x}$Ge samples with $x$ $\leq$ 0.5. The vertical dashed line indicates the $T_{\rm c}$ for $x$ = 0.5, and the solid lines in panel (c) are fits to the normal-state data by the Debye model. (d-f) Same set of data for the samples with $x$ $\geq$ 0.5. The vertical dashed line indicates the $T_{\rm c}$ for $x$ = 0.9. (g-i) Normalized electronic specific heat data plotted as a function of temperature for the samples with $x$ = 0.2, 0.5 and 0.6, respectively. In panels (a) and (b), the solid lines are entropy conserving constructions to estimate the size of specific heat jump. In panel (c), the solid curve is a fit to the data by the $\alpha$-model with $\alpha$ = 2.0.
	}
	\label{fig3}
\end{figure*}

Figure 2(a) show the temperature dependence of resistivity ($\rho$) for the samples with $x$ up to 0.5.
For $x$ $\leq$ 0.4, a $\rho$ bump is observed due to the structural phase transition.
With increasing $x$, this bump shifts to lower temperatures and becomes weakened, as seen more clearly in the inset of Fig. 2(a). At $x$ = 0.5, no such $\rho$ anomaly is discernible, suggesting that the transition is either too weak to be detected or disappears completely. This trend is corroborated by the results of magnetic susceptibility ($\chi$) measured under 4 T, as shown in Fig. 2(b). Concomitant with the $\rho$ upturn, there exist a drop in $\chi$ for $x$ $\leq$ 0.4, which is attributed to the decrease in the density of states at the Fermi level ($E_{\rm F}$).
The onsets of these two anomalies agree well and allow us to determine the structural transition temperature $T_{\rm s}$ = 190 K, 158 K, 130 K and 100 K for $x$ = 0.1, 0.2, 0.3 and 0.4, respectively.
Notably, a thermal hysteresis near $T_{\rm s}$ is present for $x$ = 0.1 [see the inset of Fig. 2(b)] but absent at higher $x$ values, signifying that structural transition is smeared out by disorder and becomes crossover like.
In the whole $x$ range of 0.1 to 0.5, the $\chi$($T$) data exhibit a weak upturn below 50 K, which can be well fitted by the Curie-Weiss law,
\begin{equation}
\chi = \chi_{0} + \frac{C}{T - \Theta},
\end{equation}
where $\chi_{0}$ is the temperature independent term, $C$ is the Curie constant and $\Theta$ is the Weiss temperature. The obtained $\chi_{0}$, $C$ and $\Theta$ fall in the ranges of 1.0$\times$10$^{-4}$ to 1.4$\times$10$^{-4}$ emu mol$^{-1}$, 0.0006 to 0.0022 emu K mol$^{-1}$, and -10 to -26 K, respectively. These $C$ values correspond to small effective magnetic moments of only 0.07-0.13 $\mu_{\rm B}$, where $\mu_{\rm B}$ is the Bohr magneton. This implies that the $\chi$($T$) upturn is most probably due to the presence of a small amount of paramagnetic impurity. The results of specific heat ($C_{\rm p}$) measurements on the $\beta$-ThRh$_{1-x}$Ir$_{x}$Ge samples are displayed in Fig. 2(c).
A $C_{\rm p}$ anomaly around $T_{\rm s}$ is detected only for $x$ $\leq$ 0.3, and it becomes smaller and broader with increasing $x$ [see the inset of Fig. 2(c)]. Except for these anomalies, the $C_{\rm p}$($T$) data look rather similar for these samples and their high temperature limit is close to the Dulong-Petit limit of 3$NR$ = 74.826 J mol$^{-1}$ K$^{-1}$, where $N$ = 3 is the number of atoms per unit cell and $R$ = 8.314 J mol$^{-1}$ K$^{-1}$ is the gas constant. Overall, these results clearly indicate that the structural transition in $\beta$-ThRh$_{1-x}$Ir$_{x}$Ge is suppressed by the increase of Ir doping and becomes no longer perceptible for $x$ $\geq$ 0.5.\\
\begin{figure*}
	\includegraphics*[width=16.5cm]{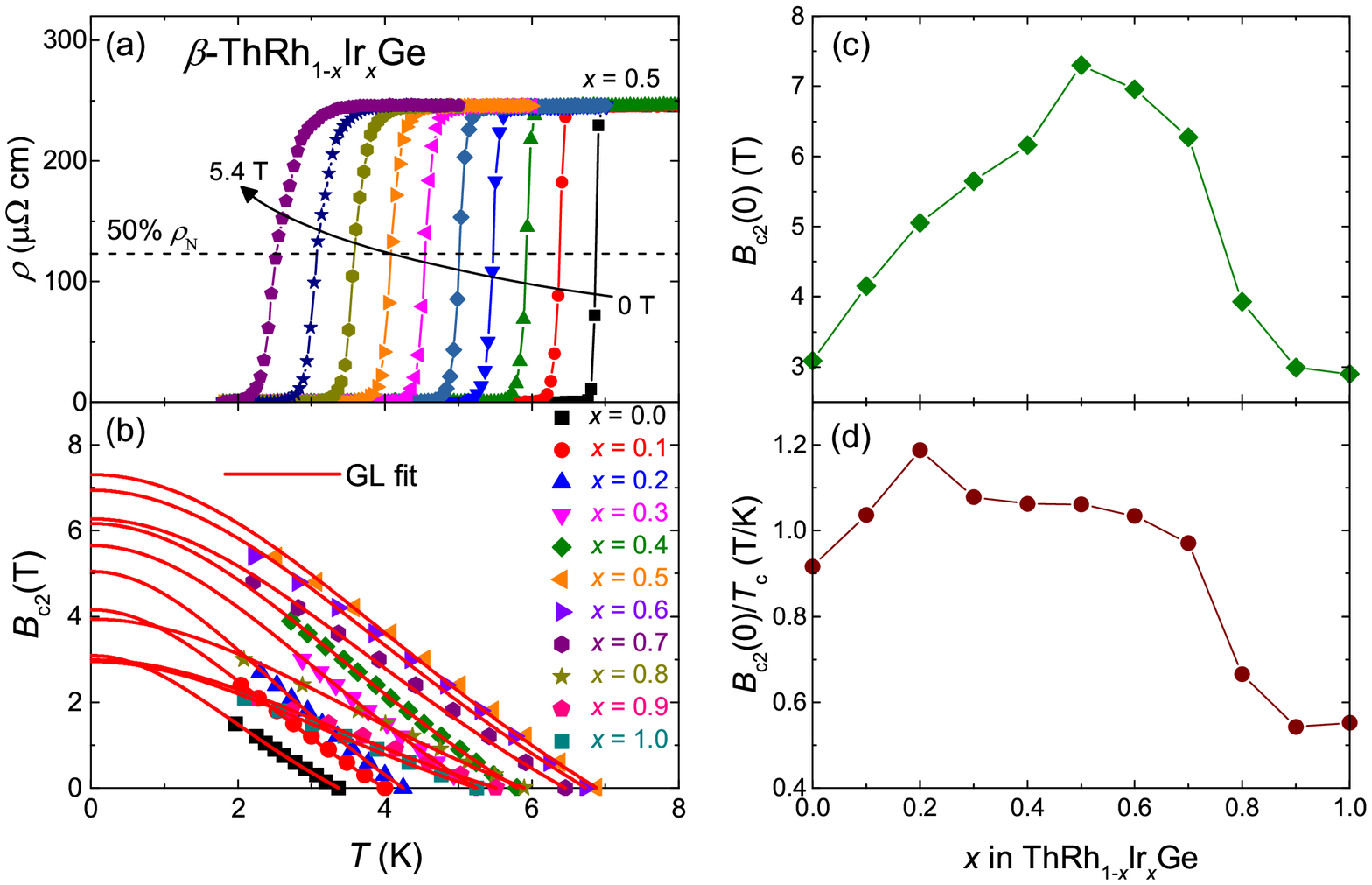}
	\caption{(Color online) \textbf{Upper critical fields in $\beta$-ThRh$_{1-x}$Ir$_{x}$Ge.}
		(a) Temperature dependence of resistivity under various magnetic fields up to 5.4 T for the $\beta$-ThRh$_{1-x}$Ir$_{x}$Ge sample with $x$ = 0.5. The dashed line denotes the midpoint of resistive transition and the arrow marks the field increasing direction. (b) Upper critical field versus temperature phase diagrams for the $\beta$-ThRh$_{1-x}$Ir$_{x}$Ge samples. The solid lines are fits to the data by GL model. (c, d) Ir content $x$ dependencies of $B_{\rm c2}$(0) and $B_{\rm c2}$(0)/$T_{\rm c}$, respectively, for $\beta$-ThRh$_{1-x}$Ir$_{x}$Ge.}
	\label{fig4}
\end{figure*}

\noindent\textbf{Evolution of superconductivity}
\begin{table*}
	\caption{Normal-state and superconducting parameters of $\beta$-ThRh$_{1-x}$Ir$_{x}$Ge.}
	\renewcommand\arraystretch{1.3}
	\begin{tabular}{p{3.0cm}<{\centering}p{1.2cm}<{\centering}p{1.2cm}<{\centering}p{1.2cm}<{\centering}p{1.2cm}<{\centering}p{1.2cm}<{\centering}p{1.2cm}<{\centering}p{1.2cm}<{\centering}p{1.2cm}<{\centering}p{1.2cm}<{\centering}p{1.2cm}<{\centering}p{1.2cm}<{\centering}p{1.2cm}<{\centering}}
		\\
		\hline
		Parameter         & $x$ = 0.0& $x$ = 0.1 & $x$ = 0.2& $x$ = 0.3& $x$ = 0.4& $x$ = 0.5& $x$ = 0.6& $x$ = 0.7& $x$ = 0.8& $x$ = 0.9 & $x$ = 1.0  \\
		\hline
		$T_{\mathrm{s}}$ (K)  & 244& 190  & 158& 130& 100& $-$& $-$& $-$& $-$& $-$ & $-$  \\
		$T_{\mathrm{c}}$ (K) & 3.36& 4.00  & 4.25& 4.63& 5.23& 6.88& 6.73& 6.46& 5.90& 5.51 & 5.25  \\
		$\gamma$ (mJ mol$^{-1}$ K$^{-2}$) & 8.09& 8.20 & 8.72& 9.48& 9.55& 12.60& 11.56& 11.20& 12.79& 11.48 & 11.80  \\
		$\delta$ (mJ mol$^{-1}$ K$^{-4}$) & 0.379& 0.449 & 0.660& 0.812& 0.879& 0.924& 0.669& 0.481& 0.317& 0.307 & 0.290   \\
		$\Theta_{\mathrm{D}}$ (K)  & 249& 235   & 207& 193& 188& 185& 206& 230& 264& 267& 275  \\
		$\lambda_{\mathrm{ep}}$      & 0.60& 0.64 & 0.68& 0.76& 0.80& 0.86& 0.81& 0.77& 0.70& 0.68 & 0.65 \\
		$B_{\rm c2}$(0) (T)     & 3.10& 4.15  & 5.05& 5.65& 6.16& 7.30& 6.96& 6.27& 3.93& 2.99& 2.90 \\
		$\xi_{\mathrm{GL}}(0)$ (nm)   & 10.3& 8.9 & 8.1& 7.6& 7.3& 6.7& 6.9& 7.3& 9.2& 10.5 & 10.7 \\
		\hline
	\end{tabular}
	\label{Table1}
\end{table*}

Figure. 3(a) shows a zoom of the low-temperature $\rho$($T$) curves for the $\beta$-ThRh$_{1-x}$Ir$_{x}$Ge samples with $x$ $\leq$ 0.5. As the increase of $x$, the resistive transition moves toward higher temperatures, indicating that superconductivity is enhanced in this doping range. It is worth noting that the transition is relatively sharp for $x$ = 0.1 and 0.5 while broadens considerably for the intermediate $x$ values. This is probably attributed to the interplay between structural transition and superconductivity, as will be discussed further below. The bulk nature of superconductivity in these samples are confirmed by the $\chi$ and $C_{\rm p}$ results shown in Figs. 3(b) and (c). Here the $\chi$ was measured under 1 mT with a zero-field cooling mode.
For each $x$ value, a large shielding fraction exceeding 110\% without demagnetization correction and a distinct $C_{\rm p}$ jump are observed.
The onsets of these anomalies coincide with the midpoint of $\rho$ drop for $x$ = 0.1, 0.2 and 0.5 (see the vertical dashed line) and the completion of resistive transition for the other $x$ values.
Hence the $T_{\rm c}$ values are determined to be 4.00 K, 4.25 K, 4.63 K, 5.23 K and 6.88 K for $x$ = 0.1, 0.2, 0.3, 0.4, and 0.5, respectively.
In the normal state, the $C_{\rm p}$($T$) data are well described by the Debye model plus electronic contribution,
\begin{equation}
C_{\rm p}/T = \gamma + \delta T^{2} + \eta T^{4},
\end{equation}
where $\gamma$ and $\delta$($\eta$) are the electronic and phonon specific-heat coefficients, respectively.
The best fits yield $\gamma$ = 8.20, 8.72, 9.48, 9.55, 12.60 mJ mol$^{-1}$ K$^{-2}$, $\delta$ = 0.449, 0.660, 0.812, 0.879, 0.924 mJ mol$^{-1}$ K$^{-4}$ for $x$ = 0.1, 0.2, 0.3, 0.4, and 0.5, respectively.
Once $\delta$ is known, the Debye temperature $\Theta_{\rm D}$ is calculated as
\begin{equation}
\Theta_{\rm D} = (12\pi^{4} N R/5\delta)^{1/3}.
\end{equation}
Then the electron-phonon coupling strength $\lambda_{\rm ep}$ can be estimated by the inverted McMillan formula \cite{PhysRev.167.331},
\begin{equation}
\lambda_{\rm ep}=\frac{1.04+\mu^{*} \ln \left(\Theta_{\rm D} / 1.45 T_{\mathrm{c}}\right)}{\left(1-0.62 \mu^{*}\right) \ln \left(\Theta_{\rm D} / 1.45 T_{\mathrm{c}}\right)-1.04},
\end{equation}
where $\mu^{\ast}$ is the Coulomb repulsion pseudopotential and assumed to be 0.13.
The resulting $\Theta_{\rm D}$ and $\lambda_{\rm ep}$ are listed in Table I.
One can see that the $\Theta_{\rm D}$ decreases from 235 K to 185 K while the $\lambda_{\rm ep}$ increases from 0.64 to 0.86 with increasing $x$ up to 0.5.
However, the further increase of Ir content $x$ above 0.5 leads to a suppression of the superconductivity in $\beta$-ThRh$_{1-x}$Ir$_{x}$Ge, which is illustrated in the Figs. 3(d-f).
Indeed, all of the $\rho$ drop, diamagnetic transition and $C_{\rm p}$ jumps move towards lower temperatures with increasing $x$ from 0.5 to 1.0.
The midpoint of the former one agree well the onset of the latter two, which gives the $T_{\rm c}$ values of 6.73, 6.46, 5.90 and 5.51 K for $x$ = 0.6, 0.7, 0.8, and 0.9, respectively.
Also, the $\gamma$, $\delta$, $\Theta_{\rm D}$ and $\lambda_{\rm ep}$ are obtained by the analysis of normal-state $C_{\rm p}$($T$) data.
As can be seen from Table I, while no systematics in $\gamma$ is found, there is an increase in $\Theta_{\rm D}$ from 185 to 267 K while a decrease in $\lambda_{\rm ep}$ from 0.86 to 0.68 with increasing $x$ in the range of 0.5-0.9, which are opposite to the trends at lower $x$ values.

By subtraction of the phonon contribution, the electronic specific heat $C_{\rm el}$ is isolated and plotted as $C_{\rm el}$/$\gamma$$T$ versus $T$ for three representative doping concentrations $x$ = 0.2, 0.5 and 0.6 in Figs. 3(g-i). It is pointed out that, for $x$ = 0.2, the monoclinic and orthorhombic phases are expected to coexist below $T_{\rm s}$.
Hence, only the normalized specific-heat jump $\Delta$$C_{\rm p}$/$\gamma$$T$ is estimated by the entropy conserving construction. This yields $\Delta$$C_{\rm p}$/$\gamma$$T$ = 1.34, close to 1.43 as predicted by the BCS theory \cite{bardeen1957theory}.
As $x$ is increased to 0.5, a rather broad $C_{\rm el}$/$\gamma$$T$ jump is observed in spite of its highest $T_{\rm c}$.
Nevertheless, the entropy conserving construction gives $\Delta$$C_{\rm p}$/$\gamma$$T$ = 1.73, which is almost 30\% larger than that at $x$ = 0.2.
This result is well reproducible among different sample batches of this doping concentration, and hence is most likely intrinsic.
Since $T_{\rm s}$ disappears around this $x$ value, the associated fluctuation may be responsible for such anomalous behavior.
In comparison, the $C_{\rm el}$/$\gamma$$T$ jump is significantly sharper at a slightly higher $x$ = 0.6.
To analyze the data, we employed the $\alpha$-model that was adapted from the single-band BCS theory \cite{johnston2013elaboration}.
This model still assumes a fully isotropic superconducting gap while allows for the variation of coupling constant $\alpha$ $\equiv$ $\Delta$(0)/$k_{\rm B}$$T_{\rm c}$, where $\Delta$(0) is the gap size at 0 K.
Note that $\alpha$ = 1.764 for the BCS theory.
For $x$ = 0.6, the $C_{\rm el}$/$\gamma$$T$ data can be well described by the $\alpha$-model with $\alpha$ = 2.0.
Given that the $T_{\rm c}$ and $\Delta$$C_{\rm p}$/$\gamma$$T$ are very similar for $x$ = 0.5 and 0.6, it is reasonable to speculate that, at optimal doping, $\beta$-ThRh$_{1-x}$Ir$_{x}$Ge behaves as a intermediately coupled, fully-gapped superconductor.
\begin{figure*}
	\includegraphics*[width=17.9cm]{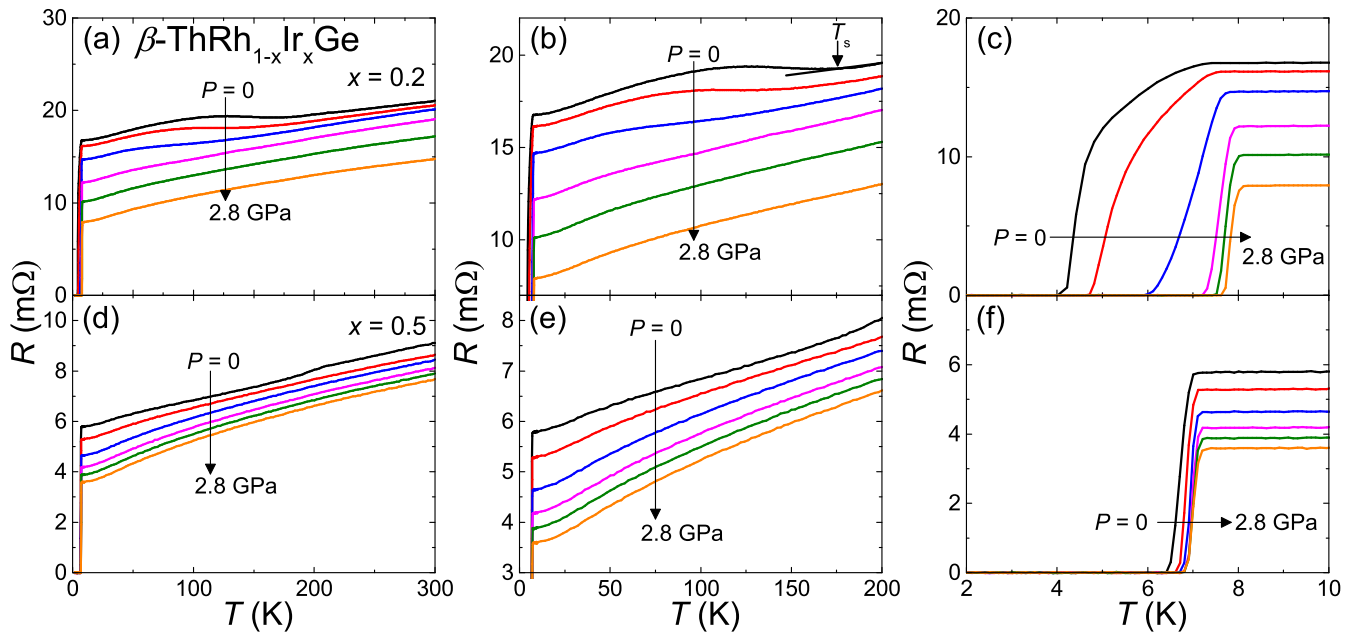}
	\caption{(Color online) \textbf{High pressure resistance data of $\beta$-ThRh$_{1-x}$Ir$_{x}$Ge at selected $x$ values.}
		(a) Temperature dependence of electrical resistance between 1.8 and 300 K under hydrostatic pressure $P$ = 0, 0.5, 1, 1.5, 2, 2.8 GPa (from top to bottom) for the $\beta$-ThRh$_{1-x}$Ir$_{x}$Ge sample with $x$ = 0.2. (b) Zoom of the data near the structural phase transition. (c) Zoom of the data near the drop to zero resistivity. (d-f) Same set of data for the sample with $x$ = 0.5.}
	\label{fig5}
\end{figure*}

The upper critical fields $B_{\rm c2}$ for all the $\beta$-ThRh$_{1-x}$Ir$_{x}$Ge samples are obtained by magnetoresistivity measurements, and an example for $x$ = 0.5 is shown in Fig. 4(a). As expected, the resistive transition gradually shifts toward lower temperatures and becomes broadened with increasing magnetic field. For each field, the $T_{\rm c}$ is determined using the same criterion as above. The resulting temperature dependencies of $B_{\rm c2}$ for all $x$ values are summarized in Fig. 4(b). The zero-temperature upper critical field $B_{\rm c2}$(0) is derived by extrapolating the $B_{\rm c2}$($T$) data to 0 K using the Ginzburg--Landau (GL) model \cite{zhu2008upper}
\begin{equation}
B_{\rm c2}(T) = B_{\rm c2}(0) \frac{1-t^{2}}{1+t^{2}},
\end{equation}
where $t$ = $T$/$T_{\rm c}$ is the reduced temperature. In all cases, the data follow nicely the GL fitting curves. In comparison, the $B_{\rm c2}$ data show upward deviation from the Werthamer-Helfand-Hohenberg model \cite{PhysRev.147.295} and the results for selected $x$ = 0.2, 0.5 and 0.6 are displayed in Supplementary Fig. S1.
The $x$ dependence of extrapolated $B_{\rm c2}(0)$ is displayed in Fig. 4(c).
One can see that the $B_{\rm c2}(0)$ achieves a maximum of 7.30 T at $x$ = 0.5, which is more than twice those of the end members.
As shown in Fig. 4(d), the $B_{\rm c2}(0)$/$T_{\rm c}$ ratio remains around 1.05 up to $x$ = 0.7, and decreases rapidly afterwards to a plateau of 0.55 at $x$ $\geq$ 0.9.
It thus appears that the pairing interaction is stronger in the Rh-rich compositions than in the Rh-poor ones.
Nonetheless, it is noted that the $B_{\rm c2}(0)$/$T_{\rm c}$ values are far below 1.86 expected for the Pauli paramagnetic limit \cite{clogston1962upper}, implying that superconductivity in the whole $\beta$-ThRh$_{1-x}$Ir$_{x}$Ge series is limited by the orbital effect.
In addition, the GL coherence length $\xi_{\rm GL}$(0) is related to $B_{\rm c2}(0)$ through the equation
\begin{equation}
\xi_{\rm GL}(0) = \sqrt{\frac{\Phi_{0}}{2\pi B_{\rm c2}(0)}},
\end{equation}
where $\Phi_{0}$ = 2.07 $\times$ 10$^{-15}$ Wb is the flux quantum.
The calculated $\xi_{\rm GL}$(0)s are listed in Table I and vary between 6.7 and 10.5 nm.\\

\noindent\textbf{Pressure effect in $x$ = 0.2 and 0.5}

Given that both the $T_{\rm s}$ and $T_{\rm c}$ of $\beta$-ThRhGe are sensitive to the application of hydrostatic pressure ($P$), it is of natural interest to investigate whether this is also the case in the Ir-doped samples. Figure 5(a) shows the temperature dependence of resistance between 1.8 and 300 K for $x$ = 0.2 under various $P$ up to 2.8 GPa. With increasing $P$, the resistance curve shifts downwards, indicating enhanced metallicity. Meanwhile, the resistivity bump due to structural transition is gradually suppressed to lower temperatures and no longer visible for $P$ $\geq$ 1.5 GPa, as seen more clearly in Fig. 5(b). On the contrary, one can see from Fig. 5(c) that the superconducting transition moves towards higher temperatures and becomes sharpened.
The overall behavior is very similar to that observed in undoped $\beta$-ThRhGe \cite{xiao2022}, and corroborates that superconductivity and structural transition are still competing with each other in lightly Ir-doped samples.
For $x$ = 0.5, whose results are displayed in Figs. 5(d-f), the increase in $P$ also leads to a decrease in normal-state resistance and an enhancement in superconductivity, although both magnitudes are much smaller compared with those at $x$ = 0.2.\\
\begin{figure*}
	\includegraphics*[width=17.9cm]{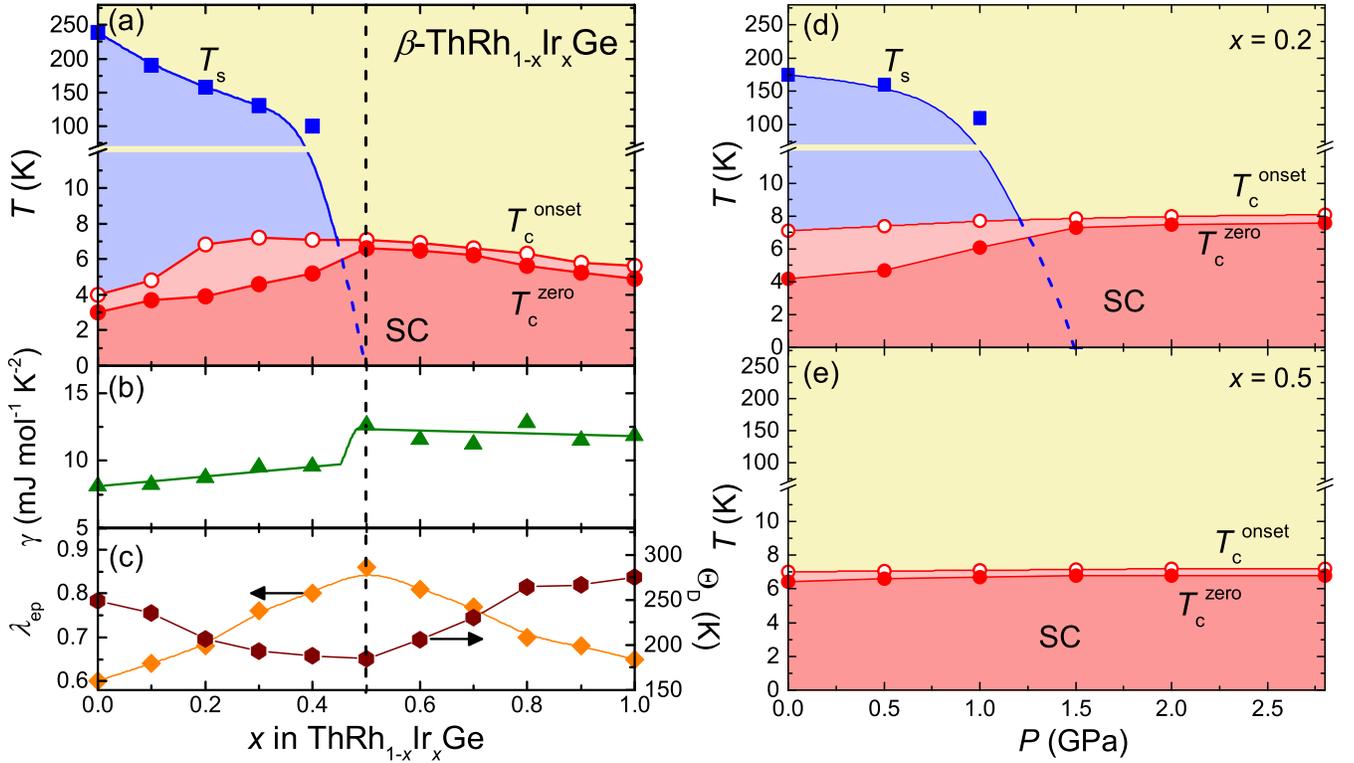}
	\caption{(Color online) \textbf{Electronic and temperature-pressure phase diagrams of $\beta$-ThRh$_{1-x}$Ir$_{x}$Ge.}
(a) $T$-$x$ electronic phase diagram of $\beta$-ThRh$_{1-x}$Ir$_{x}$Ge. Here $T_{\rm s}$, $T_{\rm c}^{\rm onset}$, $T_{\rm c}^{\rm zero}$ are the structural transition, onset and offset superconducting transition temperatures, respectively. The dashed line is a guide to the eyes.
(b-c) Ir content $x$ dependencies of $\gamma$, $\lambda_{\rm ep}$ and $\Theta_{\rm D}$, respectively, and the vertical dashed line is a guide to the eyes.
(d-e) $T$-$P$ phase diagrams for $\beta$-ThRh$_{1-x}$Ir$_{x}$Ge with $x$ = 0.2 and 0.5, respectively.
In panel (d), the dashed line is a guide to the eyes.
	}
	\label{fig6}
\end{figure*}

\begin{figure*}
	\includegraphics*[width=17cm]{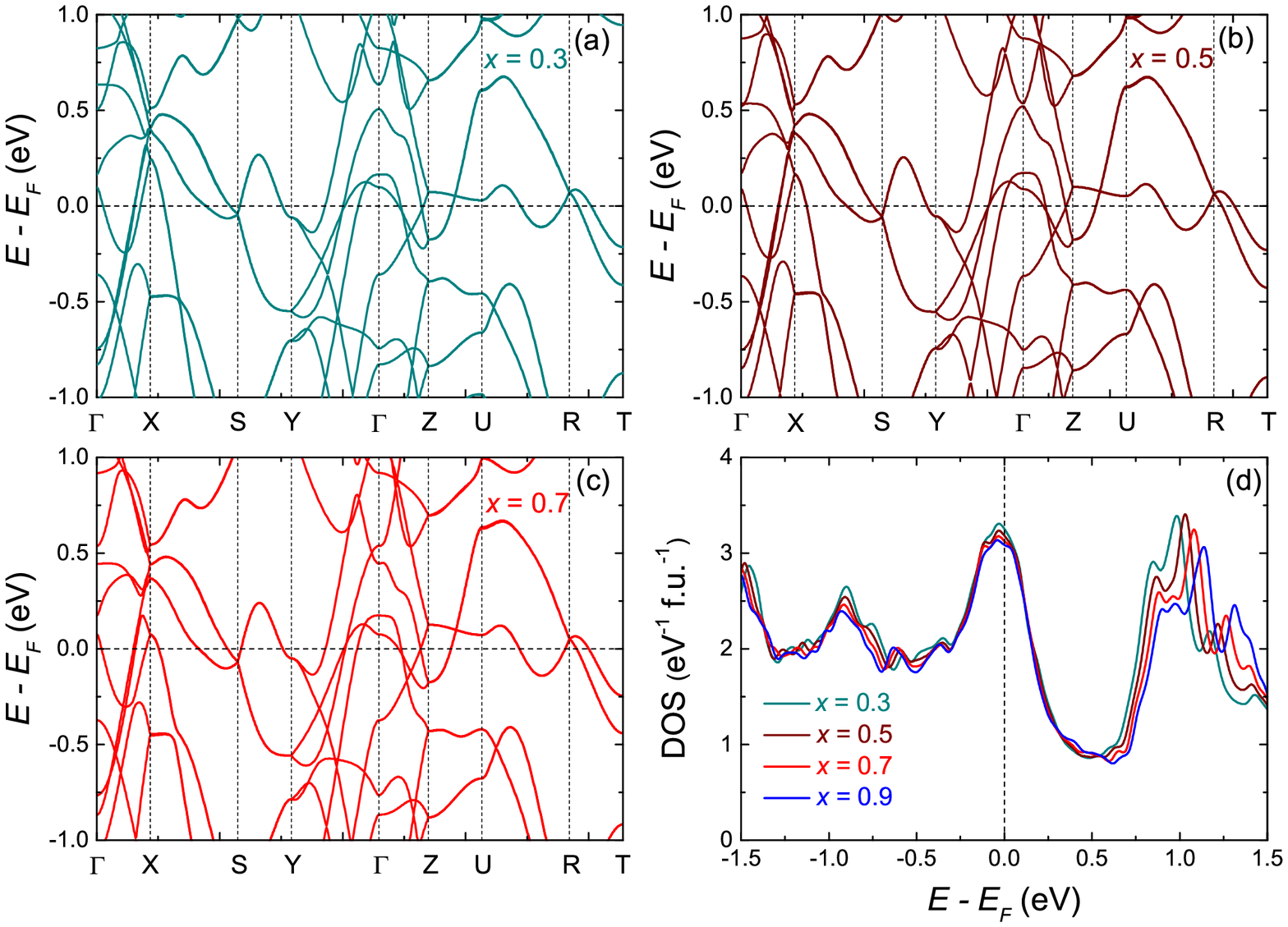}
	\caption{(Color online) \textbf{Calculated electronic band structure of $\beta$-ThRh$_{1-x}$Ir$_{x}$Ge.}
		(a-c) Electronic band structure without SOC for $\beta$-ThRh$_{1-x}$Ir$_{x}$Ge with $x$ = 0.3, 0.5 and 0.7, respectively. (d) Total density of states plotted as a function of energy for $x$ = 0.3, 0.5, 0.7, and 0.9. The vertical dashed line indicates the position of Fermi level.
	}
	\label{fig7}
\end{figure*}

\noindent\textbf{$T$-$x$ and $T$-$P$ phase diagrams}

Figure 6(a) shows the constructed $T$-$x$ phase diagrams of $\beta$-ThRh$_{1-x}$Ir$_{x}$Ge. As a consequence of Ir doping, $T_{\rm s}$ decreases monotonically and disappears at $x$ $\sim$ 0.5. On the other hand, $T_{\rm c}$ exhibits a dome-like dependence on $x$ with a maximum of 6.88 K at $x$ = 0.5, which is more than twice that (3.36 K) of $x$ = 0. It is worth noting that, while the offset $T_{\rm c}$ evolves smoothly, the onset $T_{\rm c}$ shows a step-like upturn above $x$ = 0.1 and then reaches a plateau until $x$ = 0.5. This leads to a broad resistive transition width of 2-3 K for 0.2 $\leq$ $x$ $\leq$ 0.4, as already noted above. In this $x$ range, there coexist both the orthorhombic and monoclinic phases below $T_{\rm s}$, the latter of which becomes dominant at low temperature. It is thus reasonable to speculate that the bulk superconducting transition with a lower $T_{\rm c}$ is due to the monoclinic phase, while the orthorhombic one is responsible for the onset of resistive transition with a higher $T_{\rm c}$. Indeed, as the monoclinic phase disappears at $x$ = 0.5, the difference between the onset and offset $T_{\rm c}$ is reduced strongly to $\sim$0.5 K.

To gain insight into the pairing mechanism, we plot the Ir content $x$ dependencies of $\gamma$, $\lambda_{\rm ep}$ and $\Theta_{\rm D}$ in Figs. 6(b) and (c). Two salient features are noted. First, $\gamma$ shows a modest increase with initial increasing $x$, followed by an abrupt jump at $x$ = 0.5, and tends to decrease at higher $x$ values. Second, concomitant with the $T_{\rm c}$ maximum, a maximum in $\lambda_{\rm ep}$ and a minimum in $\Theta_{\rm D}$ are observed. These features are in analogy with those observed in isovalently doped Lu(Pt$_{1-x}$Pd$_{x}$)$_{2}$In \cite{gruner2017charge}, and provides evidence for the existence of a structural QCP at $x$ $\approx$ 0.5. To further verify this scenario, the low-temperature $\rho$($T$) data are analyzed by the power law
\begin{equation}
\rho = \rho_{0} + AT^{n},
\end{equation}
where $\rho_{0}$ is the residual resistivity, $A$ is the prefactor and $n$ is the temperature exponent.
Since our samples are polycrystalline in nature, the results should be taken with caution and hence presented in Supplementary Fig. S2.
In particular, $n$ shows a minimum of $n$ $\approx$ 1.6 at $x$ = 0.5. This means a non Fermi liquid behavior due to a strong enhancement in the critical fluctuations and is a standard signature of a QCP.
It is pointed out that, in both our case and Lu(Pt$_{1-x}$Pd$_{x}$)$_{2}$In \cite{gruner2017charge}, a minimum in $n$ is observed at the QCP, indicating that they exhibit very similar behavior.
Nonetheless, the precise value of $n$ at the QCP is not universal and still the subject of debate. For example, in (Ca, Sr)$_{3}$Ir$_{4}$Sn$_{13}$, $n$ $\approx$ 1 is observed at the QCP \cite{PhysRevLett.114.097002}, which differs from those in both $\beta$-ThRh$_{1-x}$Ir$_{x}$Ge ($n$ $\approx$ 1.6) and Lu(Pt$_{1-x}$Pd$_{x}$)$_{2}$In ($n$ $\approx$ 1.8) \cite{gruner2017charge}.

We then turn the attention to the $P$-$T$ phase diagrams of $x$ = 0.2 and 0.5, which are shown in Figs. 6(d) and (e).
Intriguingly, these two diagrams look very similar with the $T$-$x$ one for $x$ in the windows of 0.2-0.6 and 0.5-0.6, respectively.
In the former case, this points to the presence of a QCP at $P$ $\approx$ 1.5 GPa, which is corroborated by a minimum $n$ $\approx$ 1.5 at this pressure obtained from the powder-law analysis of the $\rho$($T$) data under pressure (see Supplementary Fig. S3).
In comparison, the same analysis for $x$ = 0.5 indicates a slow increase in $n$ from $\sim$1.6 to $\sim$1.9 up to 2.8 GPa, substantiating that this composition is located almost exactly at a QCP.
Obviously, the roles played by isovalent Ir doping and hydrostatic pressure are essentially the same in tuning the structural transition and superconductivity in $\beta$-ThRh$_{1-x}$Ir$_{x}$Ge.
\\

\noindent\textbf{Theoretical electronic band structure}

To investigate the band structure evolution of $\beta$-ThRh$_{1-x}$Ir$_{x}$Ge, first-principles band structure calculations were performed for Ir contents of $x$ = 0.3, 0.5, 0.7, and 0.9. Given that the exact structure of the monoclinic phase remains unclear at present, all the calculations are based on the TiNiSi-type orthorhombic structure. It turns out that, in all cases, the band dispersion without considering spin-orbit coupling exhibits a strong similarity near the $E_{\rm F}$, and the results for $x$ = 0.3, 0.5 and 0.7 are shown in Figs. 7(a-c), respectively. Only at $\sim$0.5 eV above $E_{\rm F}$, a doping dependent band dispersion is discernible near the $\Gamma$ point, presumably due to the difference between the Rh and Ir states. The total DOS for $\beta$-ThRh$_{1-x}$Ir$_{x}$Ge in the $E$ $-$ $E_{\rm F}$ range of $-$1.5 to 1.5 eV is displayed in Fig. 7(d). Irrespective of the $x$ value, the $E_{\rm F}$ remains very close to a DOS peak associated with a van Hove singularity. The theoretical bare density of states $N(0)$ decreases monotonically from 3.25 to 3.08 states eV$^{-1}$ f.u.$^{-1}$ with increasing $x$ from 0.3 to 0.9. These values, together with those of $\lambda_{\rm ep}$, give the estimated $\gamma$ = 13.41, 13.94, 13.00, and 12.20 mJ mol$^{-1}$ K$^{-2}$ for $x$ = 0.3, 0.5, 0.7, and 0.9, respectively, which are in reasonably good agreement with the experimental values. The overall results support that the enhancement of superconductivity should be closely related to the phonon properties of $\beta$-ThRh$_{1-x}$Ir$_{x}$Ge, whose
evolution as a function of Ir doping is certainly of interest for future studies.

In summary, we have studied systematically the effect of isovalent Ir doping on the structural and superconductivity of $\beta$-ThRh$_{1-x}$Ir$_{x}$Ge.
The undoped $\beta$-ThRhGe displays an incomplete structural transition at $T_{\rm s}$ = 244 K and a superconducting transition at $T_{\rm c}$ = 3.36 K \cite{xiao2022}.
With increasing Ir content $x$, the structural transition is gradually smeared out and disappears at $x$ = 0.5, where $T_{\rm c}$ exhibits a maximum of 6.88 K.
This maximum is accompanied by a small jump in $\gamma$, a maximum in $\lambda_{\rm ep}$, a minimum in $\Theta_{\rm D}$ and a resistivity temperature exponent $n$ $\approx$ 1.6.
These features together unveil a structural quantum critical point located at $x$ $\approx$ 0.5, which is corroborated by high-pressure resistivity measurements.
This is supported by theoretical calculations, which indicate that the Ir doping has little effect on the electronic band dispersion near the Fermi level.
Our results suggest that $\beta$-ThRh$_{1-x}$Ir$_{x}$Ge not only offers an excellent platform to study the interplay between superconductivity and structural instability, but also provides a fresh perspective to understand the structural quantum criticality in actinide containing materials.\\

\noindent\textbf{MATERIALS AND METHODS}\\
\textbf{Sample synthesis and characterization.} Polycrystalline samples of $\beta$-ThRh$_{1-x}$Ir$_{x}$Ge were prepared by the two-step method.
Stoichiometric amounts of high-purity Th (99.5\%), Rh (99.9\%), Ir (99.9\%) and Ge (99.99\%) powders were mixed thoroughly and pressed into pellets in an argon filled glove-box. The pellets were then melted several times in an arc furnace, followed by rapid cooling on a water-chilled copper plate. The as-cast ingots were annealed in evacuated quartz tubes at 1000 $^{\circ}$C for 7 days, followed by quenching to room temperature. The phase purity of resulting samples was checked by powder X-ray diffraction (XRD) at room temperature using a Bruker D8 Advance X-ray diffractometer with Cu K$\alpha$ radiation. The lattice constants were determined by the Lebail fitting method with the JANA2006 programme \cite{petvrivcek2014crystallographic}.\\

\noindent\textbf{Physical property measurements.} Electrical resistivity and specific heat measurements were performed on regular-shaped samples in a Quantum Design Physical Property Measurement System (PPMS-9 Dynacool). The electrical resistivity was measured using a standard four-probe method and the data under hydrostatic pressure were taken in a piston-cylinder clamp-type cell using Daphne oil as the pressure transmitting medium and lead as the pressure gauge. The dc magnetization measurements were done in a Quantum Design Magnetic Property Measurement System (MPMS3).\\

\noindent\textbf{Theoretical calculations.} The first-principles band structure calculations were carried out within the density functional formalism, as implemented in the Vienna Ab-initio Simulation Package (VASP) \cite{kresse1996efficiency}. The Perdew-Burke-Ernzerhof (PBE) \cite{perdew} exchange correlation functional was used. For structural optimization, the convergence threshold of Hellmann-Feynman force and energy convergence criterion were set to 0.01 eV/{\AA} and 10$^{-6}$ eV, respectively. The wavefunction cutoff energy was fixed to 450 eV and the $\Gamma$-centered \emph{\textbf{k}} mesh was set to 5$\times$8$\times$4 in both structural optimization and self-consistent calculations. A virtual crystal approximation has been adopted to elucidate the change in the band dispersion and density of states (DOS) induced by Ir doping using WANNIERTOOLS package \cite{wu2018wanniertools}. For this purpose, two tight-binding Hamiltonians with maximally localized Wannier functions were constructed through WANNIER90 \cite{pizzi2020wannier90} for both ThRhGe and ThIrGe.

\section*{ACKNOWLEDGEMENT}
We acknowledge financial support by the foundation of Westlake University. The work at Zhejiang University is supported by the National Natural Science Foundation of China (12050003).

\section*{Competing interests}
The authors declare no competing interests.\\

\section*{Author Contributions}
G.R.X. and Z.R. conceived the project. G.R.X. synthesized the samples, did the physical property measurements and theoretical calculations with the assistance from Q.Q.Z., Y.W.C., W.Z.Y., B.Z.L., S.J.S., and G.H.C.. RZ supervised the project and wrote the paper with inputs from G.R.X..

\section*{Data availability}
The data that support the findings of this study are available from the corresponding authors upon reasonable request.

\end{document}